# Microsphere-coupled light emission control of van der Waals heterostructures

*Hyunseung Lee,[a] Van Tu Nguyen,[a,b] Ji-Yong Park[a] and Jieun Lee[a,c]\**

[a] Department of Physics and Department of Energy Systems Research, Ajou University, Suwon 16499, Korea

[b] Institute of Materials Science, Vietnam Academy of Science and Technology, Hanoi, 100000, Vietnam

[c] Department of Physics and Astronomy, Seoul National University, Seoul 08826, Korea

\* To whom correspondence should be addressed, lee.jieun@snu.ac.kr.

**Two-dimensional transition metal dichalcogenides (TMDCs) integrated to photonic structures provide an intriguing playground for the development of novel optoelectronic devices with improved performance. Here, we show the enhanced light emission from TMDC based van der Waals heterostructures through coupling with microsphere cavities. We observe cavity-induced emission enhancement of TMDC materials which varies by an order of magnitude, depending on the size of the microsphere and thickness of the supporting oxide substrate. Furthermore, we demonstrate microsphere cavity-enhanced electroluminescence of van der Waals light emitting transistor, showing the potential of 2D material based hybrid optoelectronic structures.**

**Introduction**

Semiconductor transition metal dichalcogenides (TMDCs) have recently emerged as promising candidates for photonic and optoelectronic applications due to their unique optical and electronic properties[1,2]. When thinned down to monolayers, these two-dimensional (2D) materials exhibit direct bandgaps ranging from visible to near-infrared with strong exitonic emission in contrast to their bulk counterparts[3,4]. The optical characteristics of monolayer TMDCs are widely tunable by varying thickness, doping, external fields and surrounding optical environment[5-8], making these materials excellent candidates for the development of optoelectronic devices such as light emitters, photodetectors, and optical modulators[9-12]. However, the overall quantum yield of monolayer TMDCs is reported to be low because of considerable nonradiative recombination processes[13]. It has been suggested that the quantum yield of TMDCs can be controlled by changing their optical environments, *e.g.*, through integrating these materials to optical microcavities or modulating substrate materials and geometries[12,14,15].

Integrating TMDCs to optical microcavity is reported to increase the photoluminescence (PL) emission of TMDCs through the enhancement of the local density of optical states due to the Purcell effect[16]. Recently, tailoring the PL emission of monolayer TMDCs by integrating to various types of optical cavities has been reported using Bragg reflectors, photonic crystal cavities and ring resonators[12,14,15,17-20]. Although these demonstrations incorporate sophisticated cavity design and facilitate high quality (Q) factors of cavities, complicated processes during fabrications hinder the practical application of cavity-coupling schemes. Utilizing microspheres, on the other hand, can bypass these complex fabrication processes, allowing simple and effective ways to couple TMDC monolayers to high Q microcavities. By bringing microspheres to the proximity of TMDC monolayers, it is possible to guide the emitted PL from TMDC to the surface of the cavity through the whispering gallery mode (WGM), enhancing the light extraction[21-23]. Such coupling between microsphere and TMDC monolayer has been demonstrated for monolayer MoS$_2$ directly grown on the curved surface of microspheres[24]. However, the deterministic coupling of microspheres to 2D material heterostructures has not been reported yet, which is essential for a variety of 2D van der Waals materials based photonic architectures.

In this work, we show the direct coupling of microsphere cavities on arbitrary planar monolayer TMDC and demonstrate the cavity-induced emission control. The observed emission intensity varies drastically with the cavity diameter. By increasing the supporting oxide thickness, we observe high Q cavity modes of microspheres coupled to TMDC. Furthermore, through the direct coupling of microsphere cavities to TMDC, graphene, and hexagonal boron nitride van der Waals heterostructures, we demonstrate microsphere-enhanced light extraction from electrically driven 2D light-emitting transistors. Our results show the potential of designated structures of microsphere cavities coupled to 2D materials for practical photonic and optoelectronic applications.

**Results and discussion**

Fig. 1a shows the schematics of the cavity-coupled TMDC fabricated on a planar silicon dioxide (SiO$_2$) substrate. The TMDC flakes used in our work are monolayer WSe$_2$ and MoS$_2$ which are prepared by mechanical exfoliation of bulk materials or chemical vapor deposition (CVD) synthesis method. The thickness of TMDC is confirmed by optical contrast in microscope images and photoluminescence spectra (Supplementary Fig. S1). These monolayer flakes are integrated with microspheres (diameter = 2, 5 and 7 µm) using the following method. The microspheres dispersed in water are drop casted on polydimethylsiloxane (PDMS) and dried for a few minutes. Then the microspheres on PDMS are transferred onto monolayer TMDC flakes on SiO$_2$ using a micrometer manipulator while observing with an optical microscope. During the transfer, we chose microspheres that are sparsely distributed on PDMS to obtain spatially isolated cavity systems. The final optical image of the integrated microcavity on a monolayer TMDC is shown in Fig. 1b. Our fabrication scheme is designed to deterministically couple a microsphere to a given monolayer TMDC. With this method, we could fabricate about 7 cavity-coupled systems out of 10 attempts. Such a high success rate is due to the soft adhesion of PDMS that can easily release the microspheres on targeted areas.

Using the finite-difference time-domain (FDTD) calculation packages in Lumerical, we simulated the formation of the cavity mode of microspheres, which is presented in Fig. 1c. In the simulated mode profile, the WGM of the cavity is clearly shown on the edge of the microsphere. The enhancement of the emission efficiency by the microsphere cavity is examined by calculating the quality ($Q$) factor as a function of the sphere diameter. In Fig. 1d, the calculated $Q$ is found to increase with the sphere size, suggesting a higher light extraction rate of TMDC coupled to a larger microsphere[25]. Although the calculated $Q$ values are smaller than that obtained for microspheres in air[26], we obtained a sizable $Q$ of about 265 using a 7 µm diameter sphere that is on top of the SiO$_2$/Si layer. For microsphere diameters smaller than 3 µm, the cavity mode was not able to be formed due to the weak WGM confinement.

On the fabricated TMDC-microsphere coupled system, we performed photoluminescence (PL) measurement to investigate the cavity-induced emission enhancement. For the measurement, a continuous-wave laser ($\lambda = 632.8$ nm) is focused onto the sample with an objective lens (NA = 0.6) in normal incidence. The emitted light is collected using the same objective and analyzed with the spectrometer equipped with a charge-coupled device. In the measurement, we first obtained the PL spectrum of bare WSe$_2$ without a microsphere while focusing the laser spot on WSe$_2$ layer. Then we moved the piezostage of the sample to measure the PL of WSe$_2$ coupled to a sphere. We observed up to 10 times enhancement of the PL intensity for WSe$_2$ coupled to a microsphere with the diameter of 7 µm as shown in Fig. 2a. We note that a micron thick oxide substrate was used. The enhancement rate increases as the sphere size increases from the comparison of the PL intensity normalized by the peak value of the bare WSe$_2$. Such emission enhancement of cavity-coupled WSe$_2$ was found locally only on the cavity-integrated area as verified from the spatially scanned 2D PL mapping (inset of Fig. 2a). Similar PL enhancement and the dependence on the microsphere size were observed for monolayer MoS$_2$ coupled with microspheres (Supplementary Fig. S2).

In addition to the PL intensity enhancement, we also observed the red-shift of the emission energy of monolayer TMDC as a result of the cavity coupling. To verify the origin of the red-shifting emission spectrum, we performed temperature dependent measurement and found that the observed energy shift increases at lower temperatures (Supplementary Fig. S3). The temperature dependence suggests that the asymmetry of the PL emission spectrum is induced by the exciton-phonon coupling similar to that observed in a ring-resonator-coupled WSe$_2$[27] and self-assembled quantum dots in nanocavities[28]. Although the cavity-induced PL reabsorption could also be in effect because of the overlap of the absorption line and the emission energy[29], we emphasize that our measurement shows the direct evidence of the modulated exciton-phonon coupling through the cavity integration from the observation of the systematic cavity-induced red-shift of the exciton energy as a function of temperature.

The emission efficiency enhancement due to the cavity mode is further studied by the direct measurement of the PL exciton lifetime of TMDC using time-resolved photoluminescence (TRPL). A mode-locked Ti:Sapphire femtosecond laser pulse with the central wavelength tuned at 705 nm for the excitation of monolayer WSe$_2$ is focused onto the sample. The emitted PL from the sample is measured using the time-correlated single photon counting modules. Fig. 2b shows the temporal dynamics of the PL emission of monolayer WSe$_2$ coupled to a microsphere cavity of 7 µm diameter (red dots). A faster decay time due to the cavity coupling is observed with respect to the bare WSe$_2$ (grey dots)[30]. By fitting the trace to a biexponential function, $I(t) = A\exp(-t/\tau_{Th}) + B\exp(-t/\tau_X)$, we extracted parameters relevant to the decay dynamics. Here $\tau_{Th}$ and $\tau_X$ are the thermal dissipation time of hot carriers and the exciton recombination lifetime and $A$ and $B$ are corresponding weight factors, respectively. Significant reductions of both $\tau_{Th}$ and $\tau_X$ were observed when monolayer WSe$_2$ is coupled to a microsphere cavity (details in Supplementary Fig. S4). By varying the microsphere size, the exciton recombination lifetime is modified accordingly as shown in Fig. 2d, demonstrating the Purcell enhancement of light extraction originated from the microsphere induced cavity coupling.

We have also observed that the PL intensity enhancement of cavity-coupled TMDC is largely dependent on the thickness of the oxide substrate. The systematic wet etching of the substrate was employed to vary the oxide thickness, which ranges from 100 nm to 1 µm. The PL enhancement factor was then measured for monolayer WSe$_2$ coupled to individual microspheres on different substrates for the same microsphere size. In Fig. 2c, the measured PL enhancement rate is found to increase with increasing oxide thickness. Note that in each data, the PL spectrum of cavity-coupled WSe$_2$ is normalized by the peak PL intensity of bare WSe$_2$ on the same substrate. This suggests that the mechanism of light collection using microsphere cavity is sensitive to the optical environments and the substrate geometry. Since the TRPL measurement shows the negligible dependence of the recombination lifetime on the oxide thickness (Fig. 2d), we deduce that an additional geometrical factor other than the Purcell effect is playing a role. The geometrical factor can be understood as the enhanced collection efficiency of emitted light through the combination of the microsphere with the high NA objective lens. Such lensing effect increases the light collection efficiency from TMDC for a thicker oxide substrate because of enhanced photonic jet mode[31,32], which is a critical factor when developing optoelectronic devices using microspheres.

Using the micrometer thick oxide substrate and integration to a 7 µm diameter microsphere, we then directly observed WGM of a cavity as shown in Fig. 3a. The CVD grown $MoS_2$ with the size of a few micrometers distributed uniformly on an oxide substrate was employed to fabricate TMDC-cavity coupled systems[33]. Several cavity modes with narrow linewidths are clearly observed[34,35]. Because of the flake size comparable to the contact area, the cavity modes are readily observed through the coupling of the emitted light into the microsphere. By subtracting the spectrum with that obtained on a bare $MoS_2$ near the cavity, we show the normalized spectrum of WGM in Fig. 3b. The WGM peaks indicated by blue arrows agree well with theoretical ones obtained from the FDTD simulation results. Also, the experimentally obtained cavity $Q$ was about 259 reaching the theoretical limit, showing the high quality of the integrated optical structure.

Finally, we fabricated TMDC based light-emitting transistor and demonstrated the enhancement of the cavity-coupled electroluminescence (EL) to test the feasibility of integrating microsphere cavities with van der Waals heterostructure devices. In the structure, monolayer $WSe_2$ is encapsulated by two hexagonal boron nitrides as insulating layers and two few-layer graphene as top and bottom electrodes for tunneling transistor operation[36] (Fig. 4a). The fabricated device using graphene is designed for the pure electrical control in contrast to the previously reported TMDC heterostructures[29,37]. From the electrical characterization, the onset of the tunneling current is observed at the bias voltage of about 4 V (Fig. 4b). By further increasing the bias, electrons and holes are injected into monolayer $WSe_2$ and induced EL emission as shown in the inset of Fig. 4b. The heterostructure is then integrated with a 7 µm microsphere using the method described above. On the final device, cavity-induced modification of the light extraction is examined by 2D spatial mapping of the EL spectrum (Fig. 4c). Notably, we observed about a factor of 2 local enhancement of EL intensity from $WSe_2$ coupled to the microsphere as shown in Fig. 4d. The signatures of WGMs are also observed which are indicated by blue arrows. Such enhancement was absent in the EL map obtained without the cavity as in the Supplementary Fig. S5. The observation manifests the microsphere-induced enhancement of light extraction from 2D heterostructure even without an external optical pumping. By utilizing a microsphere, we could couple multiple cavity modes within the spectral range of the emitted light, which was possible by the relatively short free spectral range of the microsphere employed in our work (~ 15 nm) compared to other cavities[38].

**Conclusion**

In summary, we have shown the PL and EL emission control of monolayer TMDC through coupling to a microsphere cavity. The PL emission intensity enhancement of about 10 times is observed using a 7 µm diameter microsphere and a micrometer thick oxide substrate. High $Q$ factors of the cavity WGM were observed through the controlled fabrication of TMDC-microsphere coupled system. The observed emission modification is explained by the combination of the cavity Purcell enhancement and the microsphere ball lensing effect. In addition, we observed cavity induced enhanced EL emission from TMDC based van der Waals heterostructure with the signature of the cavity modes, which can be further improved by optimizing the cavity design and fabrication methods.

We expect that integrating the microsphere cavity with 2D materials will be advantageous for the investigation of light-matter interactions in various regimes. For example, the coupling between the microsphere and 2D semiconductor does not require the polarization matching between the cavity mode and the emitting medium, which will allow the electron-photon interaction while maintaining the information on the electron's valley degree of freedom. The electrical pumping of the exciton emission demonstrated in our work will be potentially useful for the control of the carrier densities in 2D materials coupled with a cavity which could also lead to the observation of the exciton-polariton condensation. Moreover, recent demonstration of the chip-based fabrication of microsphere[39] shows that the scalable integration of microspheres on silicon could be possible, providing a route towards the development of the hybrid 2D photonic network through the combination of microsphere cavities, optical fibers and on-chip photonic structures.

**Acknowledgements**

The authors acknowledge supports from the National Research Foundation (NRF) of Korea (Grants No. 2020R1A2C2011334) and "Human Resources Program in Energy Technology" of the Korea Institute of Energy Technology Evaluation and Planning (KETEP), granted financial resource from the Ministry of Trade, Industry & Energy, Republic of Korea. (No. 20184030202220).

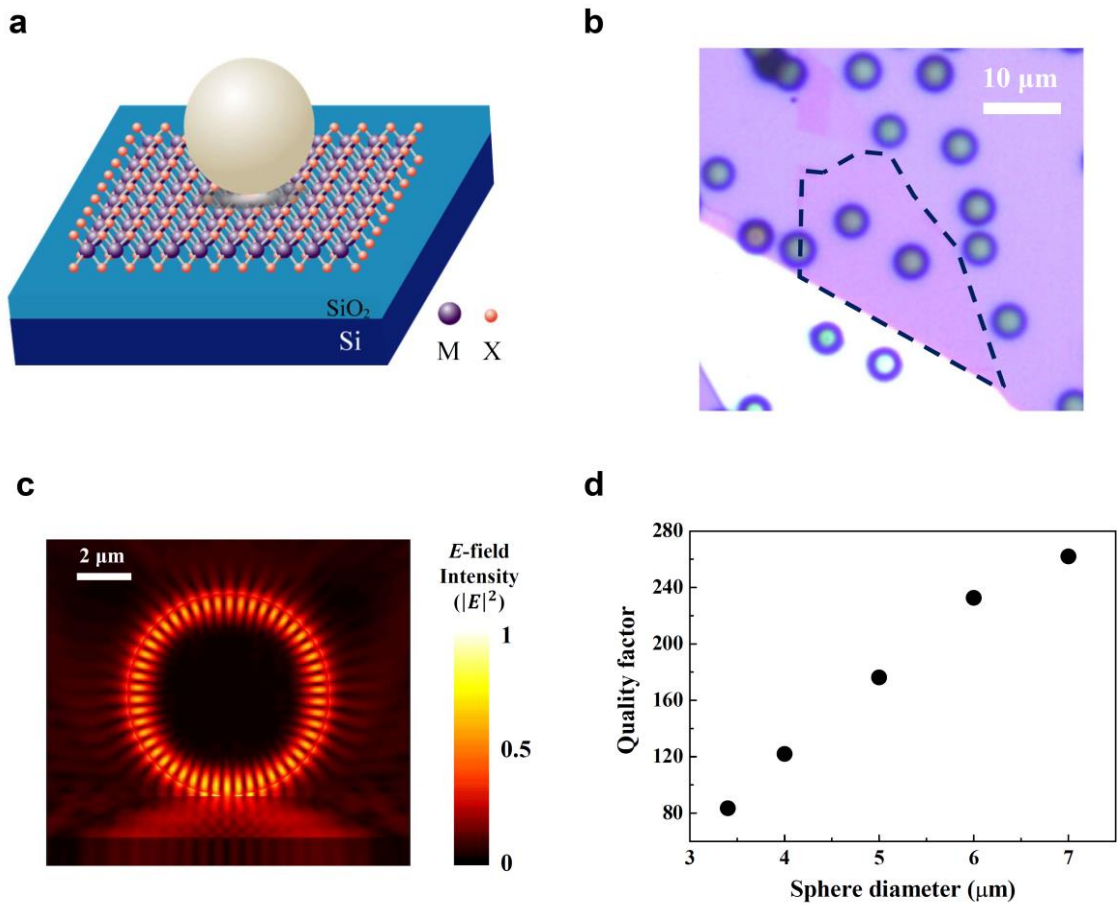

**Figure 1.** **(a)** Schematic diagram of WSe$_2$ coupled to a microsphere on SiO$_2$/Si substrate. M and X refer to transition metal and chalcogens atom, respectively. **(b)** Optical microscope image of monolayer WSe$_2$ coupled to microspheres. **(c)** Finite-difference time-domain (FDTD) simulation of the whispering gallery mode of microsphere (sphere diameter = 7 µm). Leakage mode into the substrate is observed due to the presence of an oxide layer. **(d)** Calculated Q factor as a function of the sphere diameter.

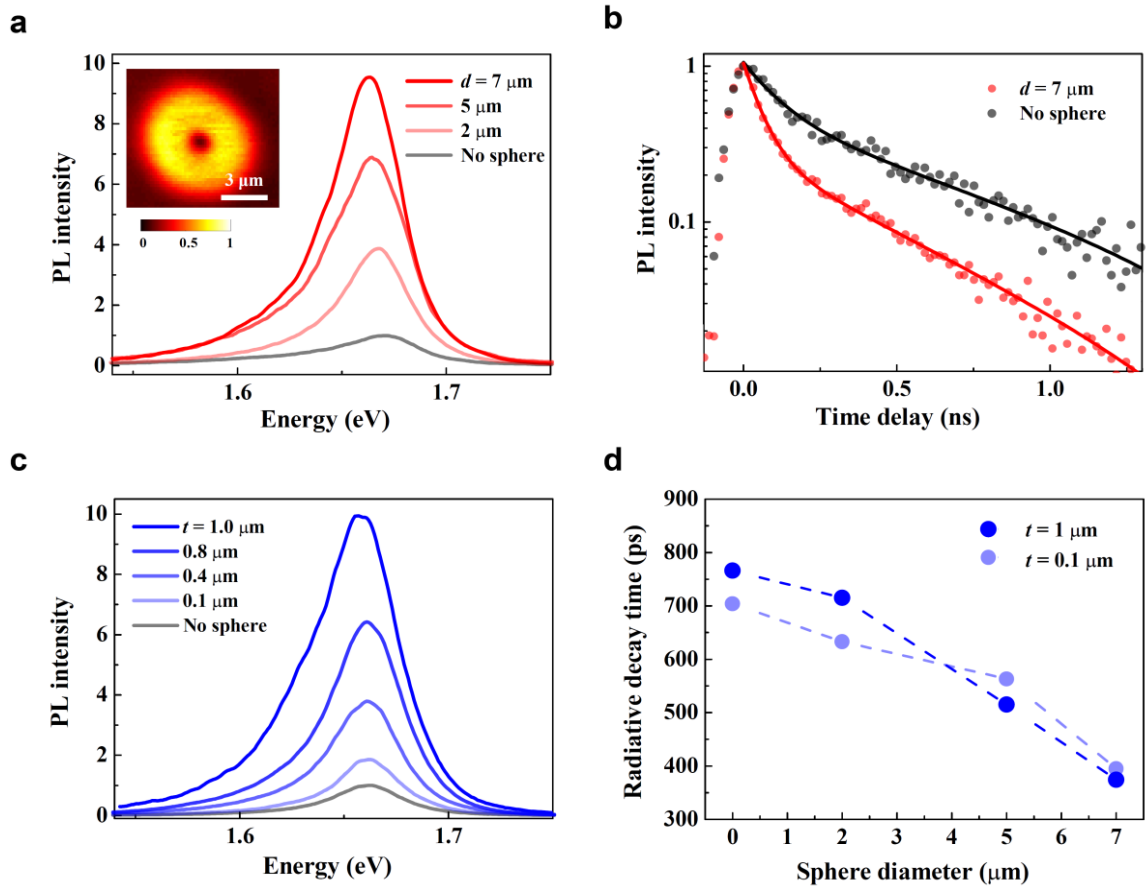

**Figure 2.** (**a**) PL spectra of WSe$_2$ coupled to microspheres with varying diameter ($d$) on a micrometer thick oxide substrate. Each spectrum is normalized by the peak intensity of the PL measured on uncoupled WSe$_2$. Inset: Spatially scanned PL image of WSe$_2$ coupled to a 7 µm diameter sphere. (**b**) Time-resolved PL of WSe$_2$ coupled to a 7 µm diameter sphere (red dots) and WSe$_2$ without a microsphere (grey dots). Fit lines using biexponential decay functions are also shown. (**c**) PL spectra of WSe$_2$ coupled to a 7 µm diameter sphere with varying oxide thickness ($t$). Each PL spectrum is normalized by the peak intensity of the PL measured on uncoupled WSe$_2$ on the same substrate. PL spectrum of uncoupled WSe$_2$ on 0.1 µm thick oxide is shown as a reference. (**d**) Radiative decay times of WSe$_2$ as a function of the sphere diameter for two different oxide thicknesses, $t = 0.1$ µm and 1 µm.

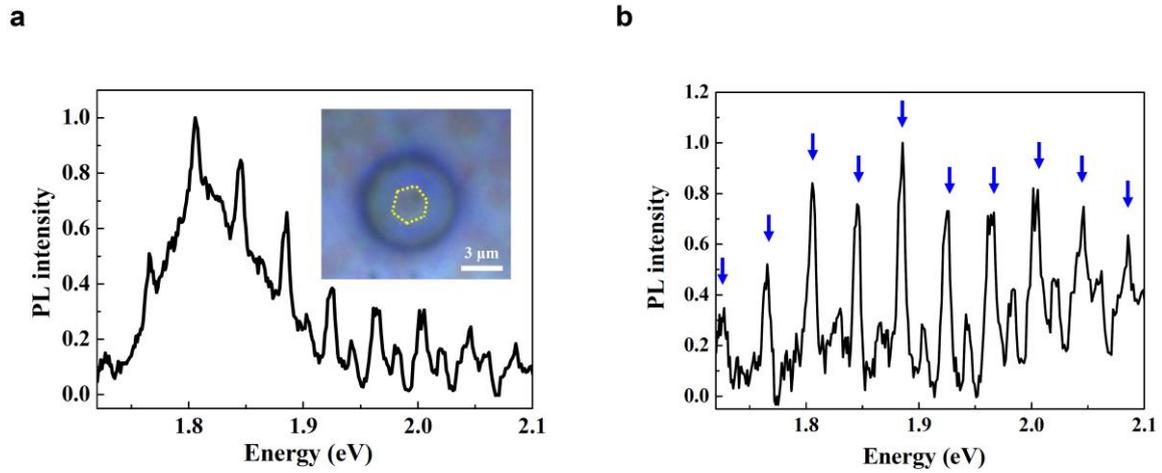

**Figure 3. (a)** PL spectrum of monolayer $MoS_2$ coupled to a 7 µm diameter microsphere on 1 µm thick oxide substrate. Inset: Optical microscope image of the CVD-grown $MoS_2$ with the integrated microsphere. The boundary of the $MoS_2$ coupled to the microsphere is indicated by yellow dashed lines. **(b)** PL spectrum of the cavity-coupled $MoS_2$ shown in **(a)** subtracted by that of the uncoupled $MoS_2$. The energies of the calculated WGMs are indicated by blue arrows.

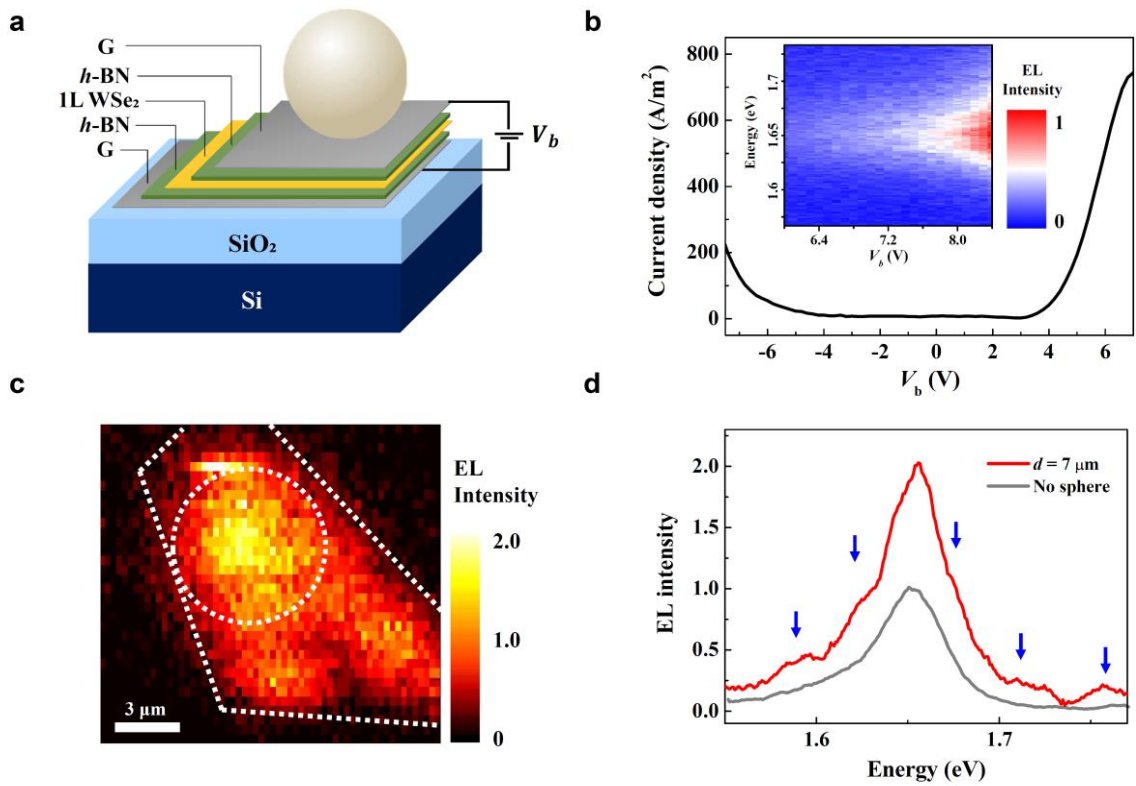

**Figure 4.** **(a)** Schematics of the microsphere-coupled van der Waals heterostructure as an efficient light-emitting transistor. Monolayer (1L) WSe$_2$ is encapsulated between two few-layer *h*-BN to avoid the direct contact with top and bottom graphene (G). **(b)** Tunneling current density as a function of the bias voltage ($V_b$). Inset: The color map of the electroluminescence (EL) as a function of the bias voltage and energy. **(c)** Spatial scan image of the EL from the cavity-coupled WSe$_2$ light-emitting transistor. Boundaries of the monolayer WSe$_2$ and the coupled microsphere ($d$ = 7 µm) are shown by white dashed lines. **(d)** EL spectra of the WSe$_2$ coupled to the microsphere (red solid line) and that without the microsphere (grey solid line). The calculated WGM energies are also shown (blue arrows).